# Interface-Assisted Room-Temperature Magnetoresistance in Cu-Phenalenyl-based Magnetic Tunnel Junctions

Neha Jha[1], Anand Paryar[2], Tahereh Sadat Parvini[1], Christian Denker[1], Pavan K. Vardhanapu[2], Gonela Vijaykumar[2], Arne Ahrens[3], Michael Seibt[3], Jagadeesh S. Moodera[4], Swadhin K. Mandal[2], Markus Münzenberg[1]

1. Institut für Physik - Universität Greifswald, Felix-Hausdorff-Straße 6, Greifswald, 17489, Germany
2. Department of Chemical Sciences, Indian Instiute of Science Education and Research (IISER), Kolkata, 741246, India
3. IV. Physikalisches Institut, Georg-August-Universitat Göttingen, Friedrich-Hund-Platz 1, Göttingen, 37077, Germany
4. Francis Bitter Magnet Laboratory and Plasma Science and Fusion Center and Department of Physics, Massachusetts Institute of Technology, Cambridge, Massachusetts 02139, United States

Delocalized carbon-based radical species with unpaired spin, such as phenalenyl (PLY) radical, opened avenues for developing multifunctional organic spintronic devices. Here we develop a novel technique based on a three-dimensional shadow mask and the in-situ deposition to fabricate PLY-, Cu-PLY-, and Zn-PLY-based organic magnetic tunnel junctions (OMTJs) with area $3 \times 8$ μm$^2$ and improved morphology. The nonlinear and weakly temperature-dependent current-voltage (I-V) characteristics in combination with the low organic barrier height suggest tunneling as the dominant transport mechanism in the structurally and dimensionally optimized OMTJs. Cu-PLY-based OMTJs, show a significant magnetoresistance up to 14% at room temperature due to the formation of hybrid states at the metal-molecule interfaces called "Spinterface", which reveals the importance of spin-dependent interfacial modification in OMTJs design. In particular, Cu-PLY OMTJs shows a stable voltage-driven resistive switching response that suggests their use as a new viable and scalable platform for building molecular-scale quantum memristors and processors.

## Introduction

Organic spintronics, a nascent field at the crossover of spintronics with organic electronics and magnetism, has attracted extensive interest and gradually developed into a new potential platform for future information technologies[1–10] due to the intriguing properties of organic materials such as long-spin coherence times[11–13], tunability of the magnetic properties by molecular design, and high chemical diversity[14]. Organic magnetic tunnel junctions (OMTJs)[15], whereby a thin layer of organic molecules is sandwiched between two ferromagnetic (FM) electrodes, have received immense attention due to their mechanical flexibility, chemically tunable electronic property, and structural fabricability[16–18]. After the first report of magnetoresistance (MR) in a organic junction[19], many groups have reported MR measurement by studying the spin injection and transport in OMTJs with different organic materials[2,5,20–24]. The performance of such devices depends not only on the properties of the organic molecule as the tunnel barrier, and magnetic injecting electrodes, but also spectacularly on the interfacial properties in the hybrid region of devices so-called "spinterface"[3,25]. This region, which arises from orbital hybridization between organic molecules and spin-split bands of the ferromagnet, can drastically influence the spin transport properties of devices. Consequently, the accurate design of the spinterface is crucial and for that some considerations must be taken into account simultaneously, i.e., the energy level alignment for facilitating the carrier injection and the spin injection/extraction, and modification of the magnetic properties of FM electrode/organic molecule interfaces[26].

Extensive investigation has been performed to improve the spinterface and hence the TMR value in OMTJs[20,27–37]. Despite getting high TMR signal in OMTJ at 2K[25] and 11K[1], however, there are very few reports at room temperature. Organic molecules are unstable at room temperature, leading to the formation of poor-quality interface between FM and organic molecules. An oxide layer of 1-2 nm is often inserted between the organics and the FM contact to enhance the MR effect[27,38–41], but their transport properties remain unclear. To explore the MR due to interfacial effect between FM and organic materials, there is a strong need to design organic molecules, which forms a stable interface up to room temperature.

Delocalized carbon-based radical molecules with an unpaired free electron, such as phenalenyl (PLY)[42,43] –based radicals, provide novel schemes for building organic spintronics devices[44], since their spin structure can be manipulated by external stimuli[42,45] (such as light, electric and magnetic fields). It has been shown that PLY coordinated with zinc ion (ZMP)-based tunnel junction generates 20% MR near room temperature[44–46].

These progresses inspired us to explore PLY and its transition metal based complex molecules-based OMTJs, i.e. PLY, Cu-PLY, and Zn-PLY. For this purpose, we developed a novel technique based on three-dimensional (3D) two-photon lithography to fabricate[47] three-dimensional mask. Low temperature (LT) ~ 80K, in-situ UHV angle deposition method combined with 3D-mask was used to construct OMTJs with an area of $3 \times 8 \ \mu m^2$. A magnetoresistance (MR) ratio $\sim 10 - 14\%$ was observed without oxide layers between FM/organic interface at room temperature, which is one of the highest MR ratios reported to date[48–50].

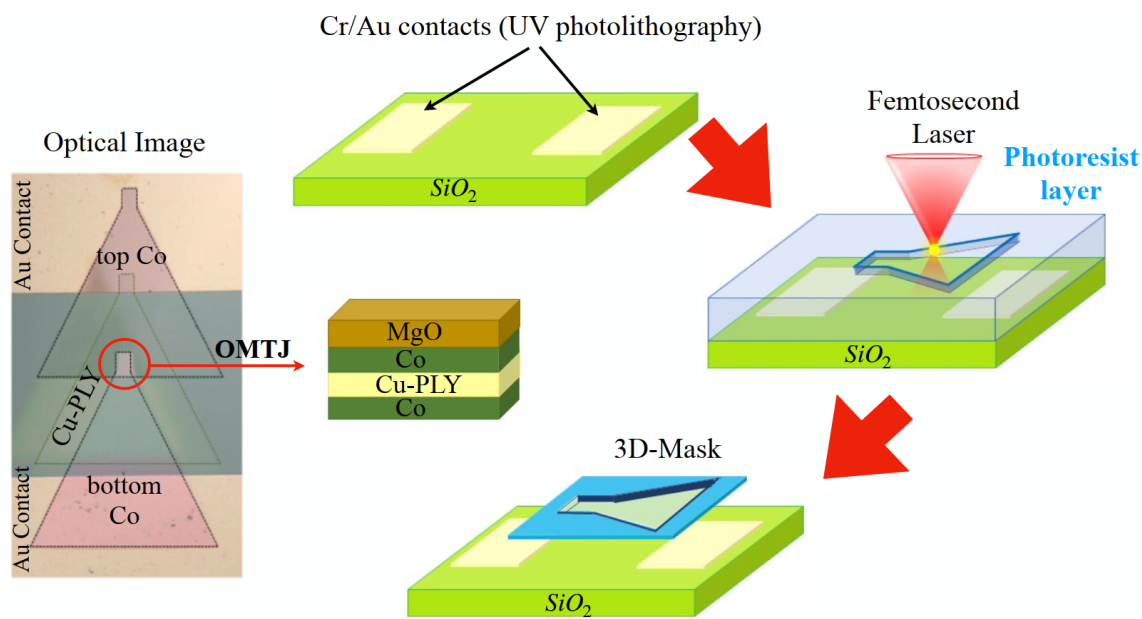

Figure1: Schematic of three-dimensional mask writing process using two-photon absorption lithography. (left) optical image of the segment used to prepare single OMTJ with top and bottom Co electrodes and resulting stack layer.

**Fabrication and optimization OMTJs**

We used two-photon absorption lithography (TPL) techniques to fabricate a 3D mask, the steps of the fabrication process are shown in Fig. 1. The use of this mask ensures precise control of the cross-sectional dimensions as well as tuning the overall size of 3D molecular junctions. Using the 3D mask, and angle deposition method as explained in detail in the supplementary material (SM Fig. S3) we fabricated multiple sets of wedge devices (SM Fig. S4) in two different categories. In the first category, wedge devices consisted of Co(8 nm)/PLY(Cu, Zn, 2-6 nm)/Co(12 nm)/MgO(4 nm), and in the second Co(8 nm)/Cu-PLY(2-6 nm)/Cu(12nm)/MgO (4 nm), which in the latter we replaced top magnetic electrodes with non-magnetic one Cu. Co/Cu electrodes and capping layers were deposited by e-beam evaporation and organic materials by thermal evaporation with an LT (~ 80K). Optical image of a single OMTJ, having two Co electrodes with an organic barrier layer is shown in Fig. 1(left).

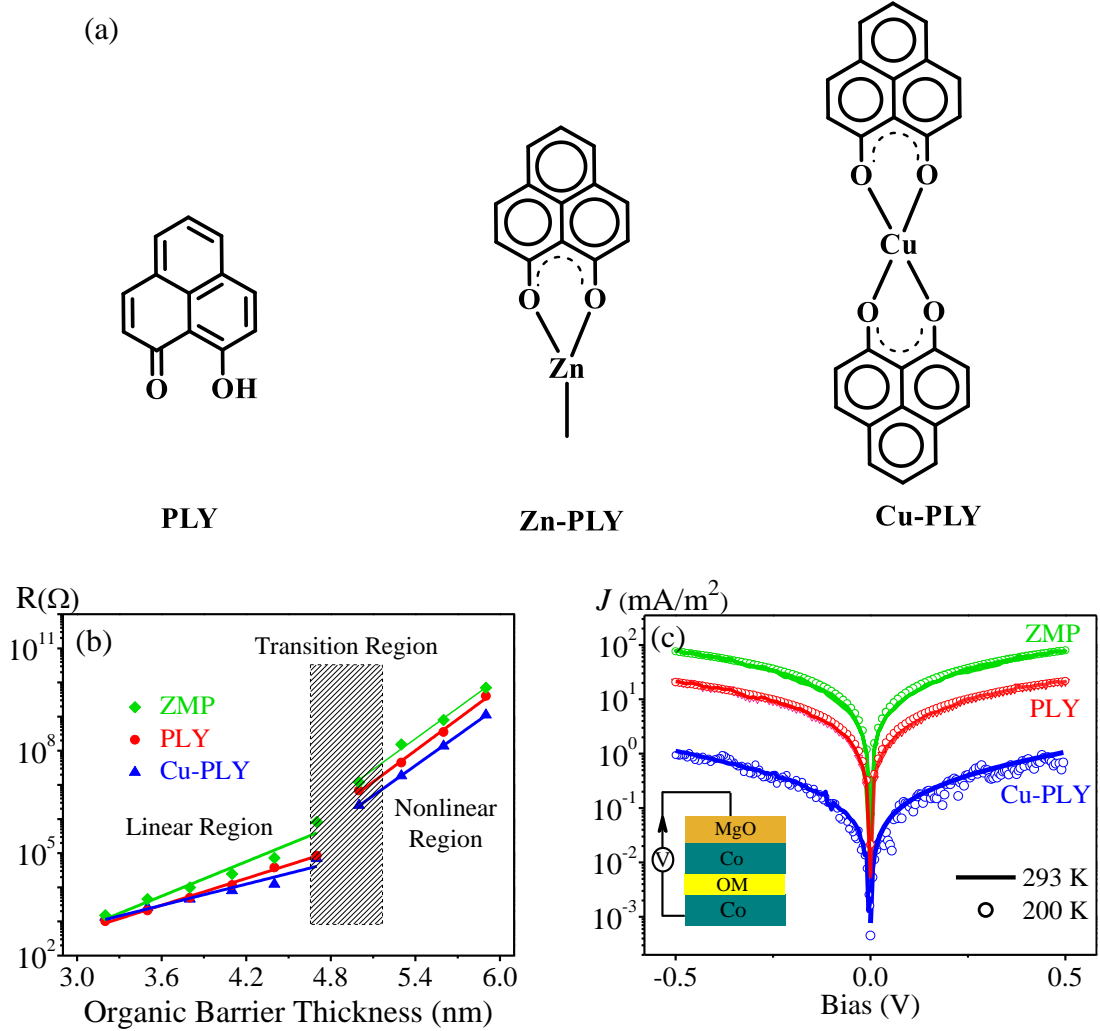

Figure 2: (a) Structure of phenalenyl and its transition metal based complex molecules. (b) Variation of the resistance as a function of the organic barrier thickness, and (c) current density versus bias voltage for OMTJs with different organic molecules at temperatures 200K and 293K.

The structure of PLY and its transition-metal-based complex molecules (Cu-PLY and ZMP) used as a barrier layer in OMTJs are shown in Fig. 2(a). To get a working TMR junction, a very thin barrier layer is required, since tunneling is expected to dominate in thin barrier junctions and decays exponentially with increasing thickness[51]. In Fig. 2(b), the resistance of the OMTJs is presented as a function of the thickness of the different organic barrier layers. Accordingly, for barrier layer thickness below 2.5 nm resistance is low (100-200 $\Omega$), because of short-circuits via pinholes or diffusion of top electrodes in the organic layer. The abrupt resistance jumps from k$\Omega$ to M$\Omega$ occurs at thickness~ 4.4 nm, afterward by increasing the thickness up to 6 nm, the resistance increases linearly in logarithmic scale from k$\Omega$ to G$\Omega$. For a thickness more than 6 nm, the resistance is in the range of 0.1 TΩ, with almost no detectable current below 1V bias. With this optimization, the organic barrier thickness of 5 nm is selected for the OMTJ device.

In order to get more insight into the origin of spin transport through organic layers, we performed current-voltage curves in the low-voltage range at 200K and 293K, using a home-built four-point probe cryostat setup. In general, the spin-dependent transport phenomenon in OMTJs is governed by tunneling and hopping mechanisms[39–41]. Inspection of data in Fig. 2(b) and Fig. 2(c) reveals that for organic-layer thickness ranging of 4.7-5.6 nm, (a) the J-V curves are clearly non-linear and symmetric, (b) resistance has a weak temperature dependence and has negligible variation at 293K and 200K, i.e. thermally activated hopping transport is ruled out (c) the J-V curves of OMTJ junctions with area A (J = I/A) are well-matched with the theoretical Brinkman model based on the Wentzel− Kramers−Brillouin (WKB) approximation for estimation of barrier height "φ"(eV) and tunneling barrier thickness "d" (nm) (details see in SM table 1). These results prove well that, tunneling is the dominant transport mechanism in the optimum OMTJs.

## Tunneling magnetoresistance (TMR) of Cu-PLY-based OMTJs

In this section, we focus on the magnetoresistance (MR) of OMTJs with the Cu-PLY molecule as the barrier layer. The magnetoresistance in such tunnel devices is referred to as tunneling magnetoresistance (TMR), which arises from the magnetization-dependent total density of states of the tunneling electrons at the surface[45]. The Transmission electron microscopy (TEM) images of Cu-PLY-based OMTJs show that all the interfaces are sharp and without interdiffusion, see Fig. 3(a). The nonlinear current-voltage (I–V) characteristics with φ =1.32 eV suggest tunneling as the dominant transport mechanism (see Fig. 3(b)). Moreover, the magnetic properties of the OMTJ were characterized using magneto-optical Kerr effect (MOKE) measurement technique at room temperature. The hysteresis loops for the top and bottom electrodes of the device in Fig. 3(c) shows coercive field of $H_c$~ 10 mT and $H_c$~ 15 mT, respectively, although as we showed in SM Fig. S5, for 8 nm Co thin film $H_c$~ 6 mT. The magnetic moment/coercive field modification depend strongly on the hybridization between the π-electrons of the Cu-PLY molecular layers and the 3d-band electrons of the Co metal. Furthermore, as a result of the hybridization, the electronic states at the interface change, leading to a new hybrid interface called "Spinterface" [3,25].

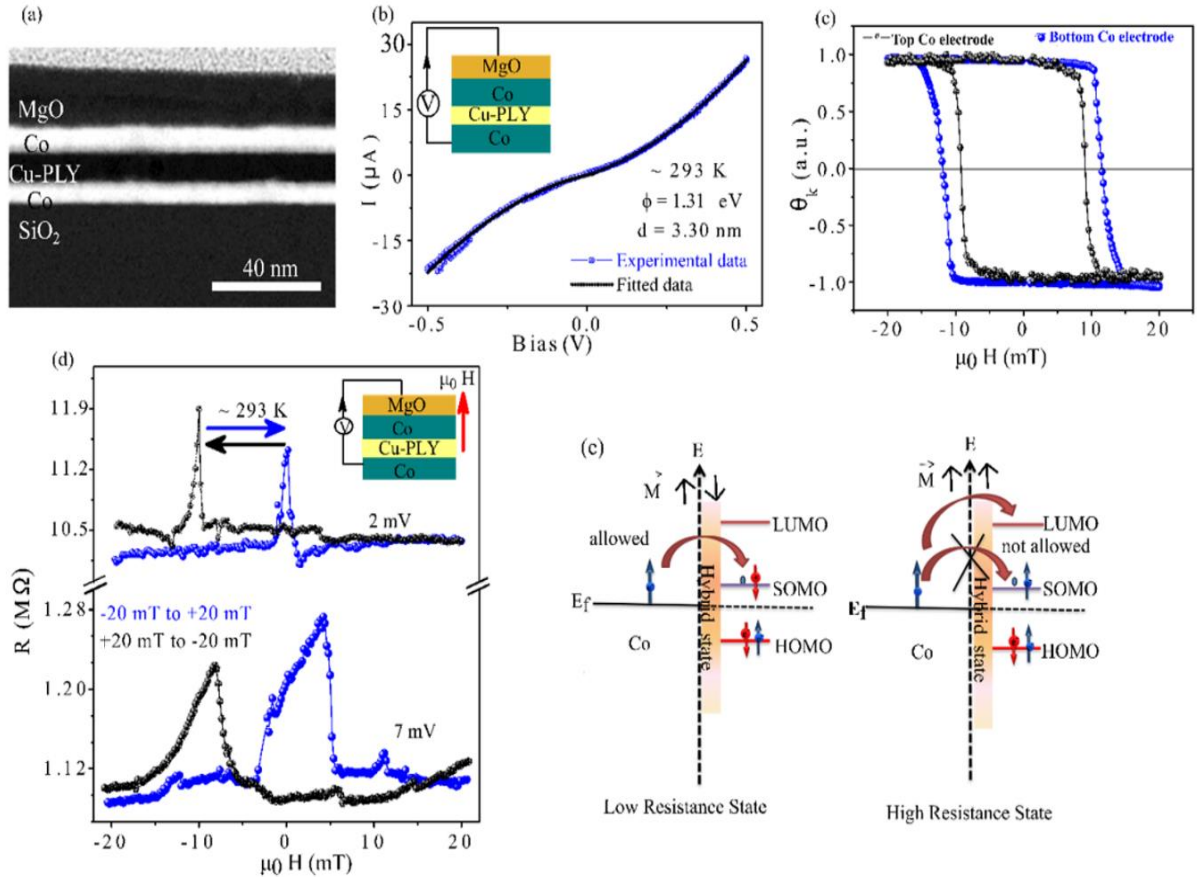

Figure 3: (a) TEM image of junction showing sharp interface without any diffusion of interlayer. (b) Non-linear I-V at room temperature. (c) Hysteresis loop of the OMTJ shows a switching field near 10 mT. (d) MR measurement with two different applied bias voltages, top for 2 mV and bottom for 7 mV. (e) Model for explaining for MR mechanism,

Variation of the resistance as a function of the in-plane magnetic field Amplitude is shown in Fig. 3(d). The TMR can be defined as TMR = $\frac{R_{AP}-R_P}{R_P} \times 100$, where $R_P$ and $R_{AP}$ are the tunneling resistance when magnetization of the two electrodes are aligned in parallel (p) and antiparallel (AP), respectively. We found TMR ~ 14% for Cu-PLY-based OMTJ, which lies among the highest value obtained until now in a device without adding any oxide tunnel barrier between FM/organic interface at room temperature. This is a very promising result since the reports of room temperature TMR effects in FM/organic/FM devices are still very limited in organic spintronics.

To understand the mechanism behind TMR signal with Cu-PLY OMTJ, we used a model similar to Raman et.al model[53]. The offset between the nearest molecular orbitals and the electrode Fermi determines the tunneling barriers for OMTJs. It allows identifying the main charge carriers flowing through the device. Fig. 3(e) shows schematically the Co Fermi level and the lowest unoccupied molecular orbitals (LUMO), highest occupied molecular orbitals (HOMO), and semi occupied molecular orbitals (SOMO) of radical state of the Cu-PLY molecule. As the HOMO is fully occupied, the SOMO of the molecule with a free electron with spin up or down participates in transport. It is very well established that the SOMO of the odd alternant hydrocarbon phenalenyl-based radical is formally a nonbonding molecular orbital (NBMO) and hence during transport it experiences a minimal change in reorganization energy[54]. The electron tunneling from the metal to the molecular orbitals is channeled through SOMO and depends on the electron spin direction. The model (in Fig 3(e)) we discuss in the following is based on Raman et al.[44]. Due to the strong exchange bias coupling of the Cu-PLY molecule to Co surface a new hybrid interface is formed. The origin of the MR signal can be explained based on the spin selective transport through the hybrid interface. When the spin of Co and the SOMO level are aligned anti-parallel (Fig. 3(e) left), the spin-up electron from metal can be injected into the first molecule layer, as finds an empty state in the SOMO level, which leads to lower resistance. On the other hand, when the FM layer and the interface layer are aligned parallel (Fig.3(e) right), the spin of Co cannot find an empty state in the SOMO level of the Cu-PLY molecule, hence the spin tunnel to LUMO level, which needs high energy and high resistance state. Consequently, the Cu-PLY layer can be a spin filter with a high MR effect at room temperature.

In the following, we replace the top magnetic electrode with a nonmagnet Cu electrode and investigate MR. As shown in Fig. 4(a), the interface switching namely the "Interface magnetoresistance (IMR)" effect was observed. Independent switching of the bottom Co with respect to an interface layer is proposed to give rise to the Interface magnetoresistance (IMR) effect, due to the strong $\pi$-d hybridization between molecular $\pi$ −orbitals and the Co d-states which alter magnetic properties of Co surface[3]. IMR ratio IMR $= \frac{R_{AP}-R_P}{R_P} \times 100 \sim 10\%$, where $R_P$ (as see in fig 5(a), $R_P = \frac{R_{P1}+R_{P2}}{2}$) and $R_{AP}$ are the resistance when magnetization of Co electrode and interface are aligned in P and AP states, respectively.

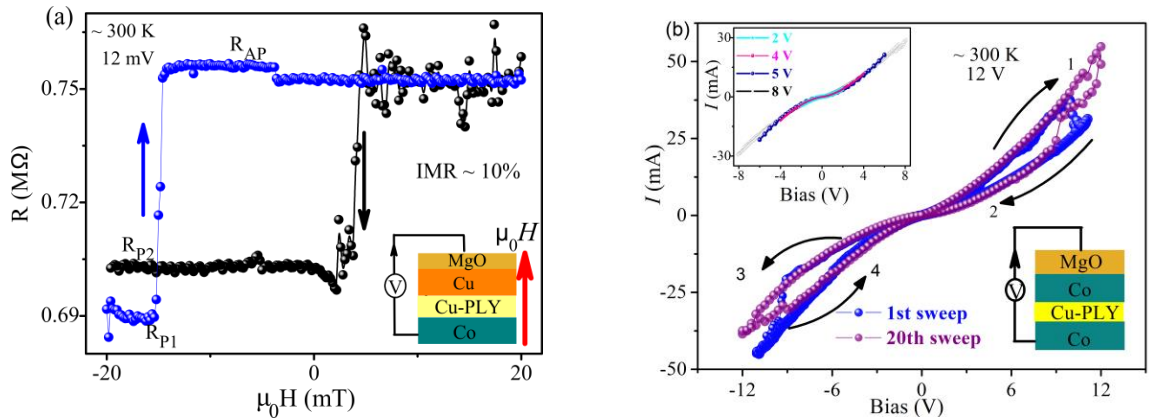

Figure 4: (a) Resistance variation with applied magnetic field for OMTJ Co/Cu-PLY/Cu/MgO at room temperature showing IMR effect. (b) Current–voltage (I-V) room temperature characteristics at H = 0. The irreversibility is clearly visible at both positive and negative voltages (bias). Inset: I-V characteristics from 3 V to 8 V reflect that at low voltage range there is no hysteresis appeared.

It is worth mentioning that by controlling the effects that appear at higher bias voltages, and we will discuss further, higher IMR is accessible. We found evidence that multiple 100% IMR could potentially be realized, consistent with spinterface owning a spin blockade effect.

Additionally, with the MR effect, as shown in Fig. 4(b), the current-voltage (I–V) characteristics of Co/Cu-PLY/Co junction demonstrate a novel preliminary electric memory effect. By increasing the bias voltage from zero to positive voltages, it can be seen that the device is in a low conductivity state with less than half of the resistance. Current sharply increases from 20 mA to 50 mA. Within a hysteresis, when the voltage is decreased back to zero, the device returns to the high conductance state. A similar process occurs at negative voltages. This I-V behavior at high bias, indicates that the device resistance can increase or decrease depending on the polarity of bias voltage. This hysteresis behavior exhibits a typical "pinch-off"[55] feature like in memristive devices similarly shown in[56], operated at different voltage levels with different current responses. The continuous multiple sweep (for 20 cycle) measurements return no significant degradation in the performance of the OMTJs, which indicates their stable behavior. Furthermore, voltage-driven resistance switching phenomena has several mechanisms in different molecular spintronic devices[57–59], but the most dominant is the trapping of charges on the molecules. In this case, OMTJ based memristive devices change their resistance by varying the direction of the spin of the electron. Not only Cu-Ply but also PLY and ZMP show the same effect (see SM Fig.S8). The final device magnetization state is determined by the accumulative effect of electrons and spin excitations. Thus, device conductance/current depends also on the integral effects of the I-V profile. For the device that means the MR effect is always connected to memory effects in molecular layer.

## Conclusion

We used the elegant approach of in-situ deposition technique with a home-built 3D mask to fabricate PLY, Cu-PLY, and Zn-PLY based OMTJs at ~ 80K, without adding any additional interface separation. The optimal tunnel barrier thickness of ~ 5 nm was identified by comparing current-voltage response and resistance as a function of barrier thicknesses. Our studies have shown tunneling as the dominant electron/spin transport mechanism in optimized OMTJs. We find room-temperature TMR ~ 14% for OMTJs with Co bottom and top electrodes, which lies among the highest values reported so far in devices without oxide layer, and IMR~ 10 % for OMTJs with a Co bottom electrode and Cu top electrode. The effects are found to be related to the formation of specific hybrid layer at the Co/Cu-PLY interface. The results demonstrate the potential for controlling parameters such as the sign and magnitude of magnetoresistance by engineering the interfacial properties. Especially, at high bias in absence of magnetic field, we observed voltage-driven resistive switching like memristive behavior for all three molecules based OMTJs. The device is highly stable at high bias and the corresponding hysteresis I-V curves are smooth without exhibiting any abrupt change in current after the 20th cycle. These finding, pave the way for the development of room temperature molecular memory devices and memristors.

## Acknowledgments

This work was supported by the Landesgraduiertenfördung Mecklenburg-Vorpommern and by the Deutsche Forschungsgemeinschaft (DFG, German Research Foundation). N.J gratefully acknowledges funding by the LGF. I would like to acknowledge Dr. Alessandro Lodesani for fruitful discussion.

Contribution

M.M and S.K.M designed the original research approach of this work. M.M, C.D and N.J conceived the idea of study spin dependent transport on OMTJ. C.D and N.J design the experiment and fabrication of devices with 3D mask technology. N.J carried out the experiments and analyzed the data. A.P synthesized the Cu-PLY molecule, VG characterized the Cu-PLY molecule by single crystal X-ray, P.K prepared PLY, Zn-PLY and Cu-PLY molecules. S.K.M supervised the molecular materials preparation and characterization. Arne carried out TEM measurements. All authors contributed to writing the manuscript.

All authors declared no competing interest.

**Supplemental Material: Interface-Assisted Room-Temperature Magnetoresistance in Cu-Phenalenyl-based Magnetic Tunnel Junctions**

Synthesis and Characterization of the Cu-PLY Molecule

Synthesis process of $Cu^{II}(PLY)_2$ refer as Cu-PLY molecule is shown in Figure. S1. The molecule's synthesis and initial characterization like cyclic voltammetry (CV) and Electron paramagnetic resonance (EPR) spectroscopy are performed in Prof. Swadhin Mandal's laboratory IISER Kolkata, India .

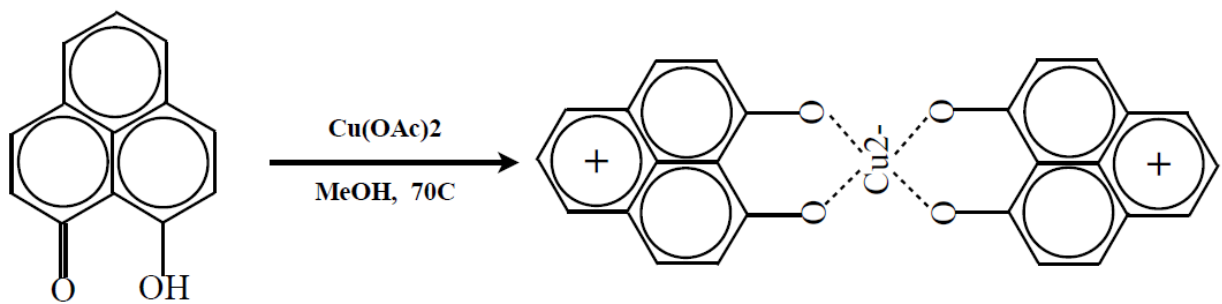

Figure S1: Schematic of synthesis of the Cu-PLY molecule from 9-hydroxy phenalenone[1].

The ligand 9-hydroxy phenalenone (PLY) is prepared from literature method used in. PLY (1, 0.98 g, 5 mmol) is dissolved in 60 mL Acetonitrile (ACN) and heat at 60°C to dissolve the ligand. To this clear hot solution, mathanplic (MeOH) solution (30 mL) of $Cu(OAC)_2 H_2 O$ (0.454 g, 2.5 mL) add dropwise. Immediately, the yellow-brown precipitate is formed. The reaction mixture is heated at 60°C for another 3h with vigorous stirring. Then the content is allowed to cool at room temperature and the precipitate is filtered. The precipitate is washed repeatedly with ACN and MeOH to remove any unreacted ligand and metal salt. The solid is vacuum dried and then collected. Crystals suitable for SCXRD are grown by dissolving vacuum dried complex in dimethylformamide (DMF) at room temperature with yield of 79% shown in. Analytically calculate for $C_{26} H_{14} CuO_4$, C: 68.79, H: 3.11; Found: C: 68.78, H: 3.36.

The cyclic voltammetry (CV) of $Cu^{II}(PLY)_2$ molecule with two irreversible reduction states, was carried out using Perkin Elmer 2400 series II, CHNS/O analyzer under argon atmosphere (Figure. S2(a)). Two reduction peaks at -1.256 V and 1.449 V indicate that Cu-PLY can exhibits reduction to radical state and di radical Cu(II) complex state. Furthermore, an intense peak at -0.067 V seen found to be responsible for the reduction process.

The room-temperature EPR spectrum (Figure. S2(b)) centered at ~ 310 mT exhibits a relatively broad absorption response with two g-factor, i.e., $g^{\parallel}$ = 2.07 and, $g^{\perp}$= 2.17. Since the g-factor of a free electron is 2.0023, it is implied that the angular momentum of the Cu (II) ion in this molecule has been altered through SOI due to the presence of an unpaired electron in the PLY radicals state. The EPR absorption can be attributed to mono-nuclear Cu (II) species.

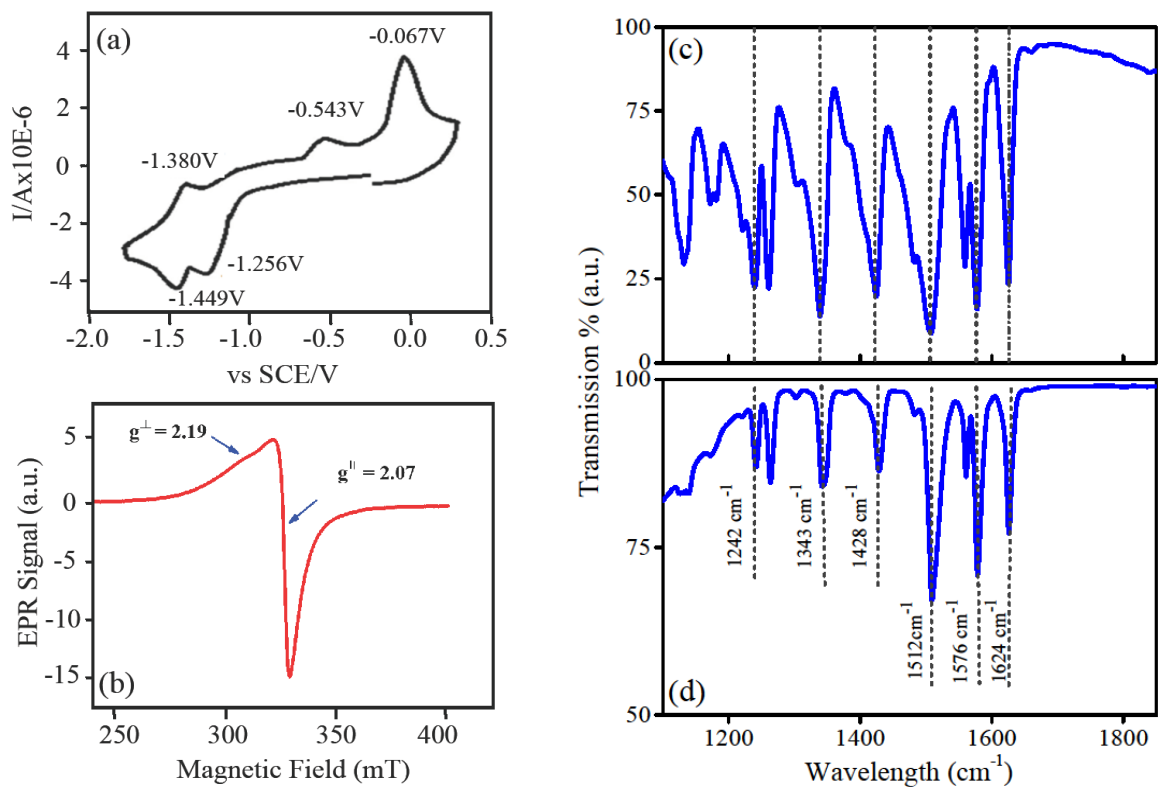

Figure S2: (a) Cyclic voltammetry of Cu-PLY molecule (100 mM TBAP) in dry dimethylformamide (DMF) under argon Pt as working, supporting and reference electrode and referenced to Fc/Fc+. (b) Room temperature X-band EPR spectrum of $Cu^{II}(PLY)_2$ powder. (c) FTIR spectra of Cu-PLY powder in KBr pellet and (d) its thin film deposited on $Si/SiO_2$ substrate.

The Cu-PLY molecular thin film was grown on a commercially available 500 nm $SiO_2$ substrate by thermal evaporation technique. Insight into the interaction of the substrate with molecule as well as thin-film quality can be inferred from Fourier-transform infrared spectroscopy (FTIR). The same range (1200-1800 $cm^{-1}$) and position of peaks in FTIR spectra of the Cu-PLY molecule thin film (Figure. S2(c)) and bulk (Figure. S2(d)), reveal that molecules or molecular bonds are not damaged during deposition on $SiO_2$. When comparing the spectra of the powder and the thin film Figure. S2(c) and (d), the presence of functional groups such as an aromatic and conjugated alkanes confirms the aromaticity of the ring system in Cu-PLY is maintained even during thermal deposition of Cu-PLY. The region from 1200 -1395 cm$^{-1}$ resulting weak to moderate stretching and bending of CO, CC, and CH bond[2] absorption peak. Despite the peak positions are identical but there are slight differences in the transmission intensity, which could be expected due to the possibility of the modification of lattices and crystal structure of Cu-PLY molecules powder due to thermal heating (more information is given in molecular supplementary SI-2).

## OMTJ Devices Fabrication and characterizations

In this study, we used a three-dimensional (3D) two-photon lithography technique to fabricate molecular tunnel Junctions with high-performance[3]. This technique can ensure precise control of the cross-sectional dimensions as well as tuning the overall size of the 3D structures.

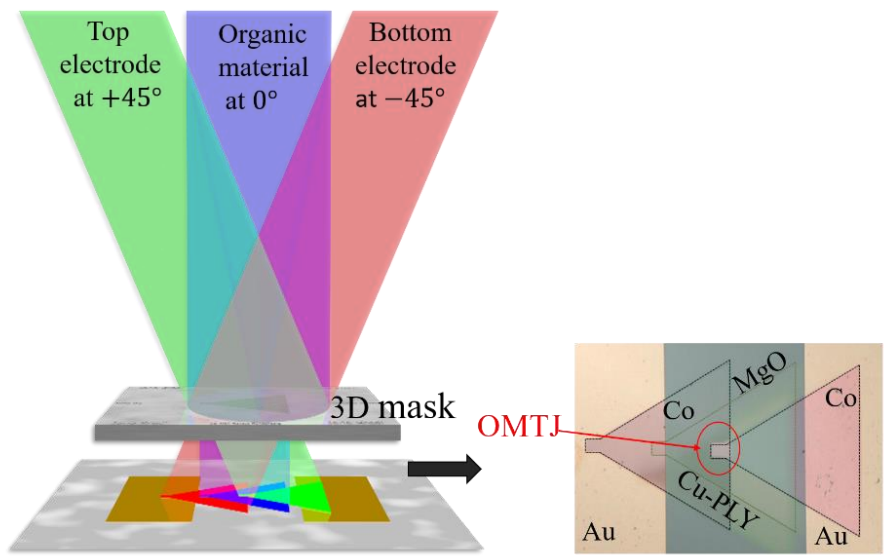

Figure S3: Schematic representation of the in-situ angle deposition of OMTJ devices.

For this purpose, first the Cr (10 nm)/Au (100 nm) gold contact pads were fabricated using a standard lift-off photo-lithography process with positive resist (Allresist@ AR-P 3840). Then, the commercial 500 nm SiO$_2$ substrate was adhered to the sample holder of direct laser writing (DLW) setup with a droplet of IP-S resist on the top side. The laser power used during 3D mask writing is 30 mW with a scan speed of 6000 µm. Using LN2 the substrate is cooled down to 80 K in the thermal evaporation chamber and transfer to the e-beam evaporation chamber to make bottom Co electrode at $-45°$ angle. Organic layer at $0°$ is deposited by thermal evaporation with deposition rate of 0.01 nm/sec. Finally top electrode at $+45°$ angle and the capping layer of MgO (4 nm) at $0°$ are deposited by e-beam evaporation (Figure. S4). Since, the thickness optimization of the organic barrier is crucial for the device performance, wedge shape structured OMTJ devices are fabricated by fixing the thickness of top and bottom Co electrodes and varying the Cu-PLY molecule thickness (Figure. S5(a)).

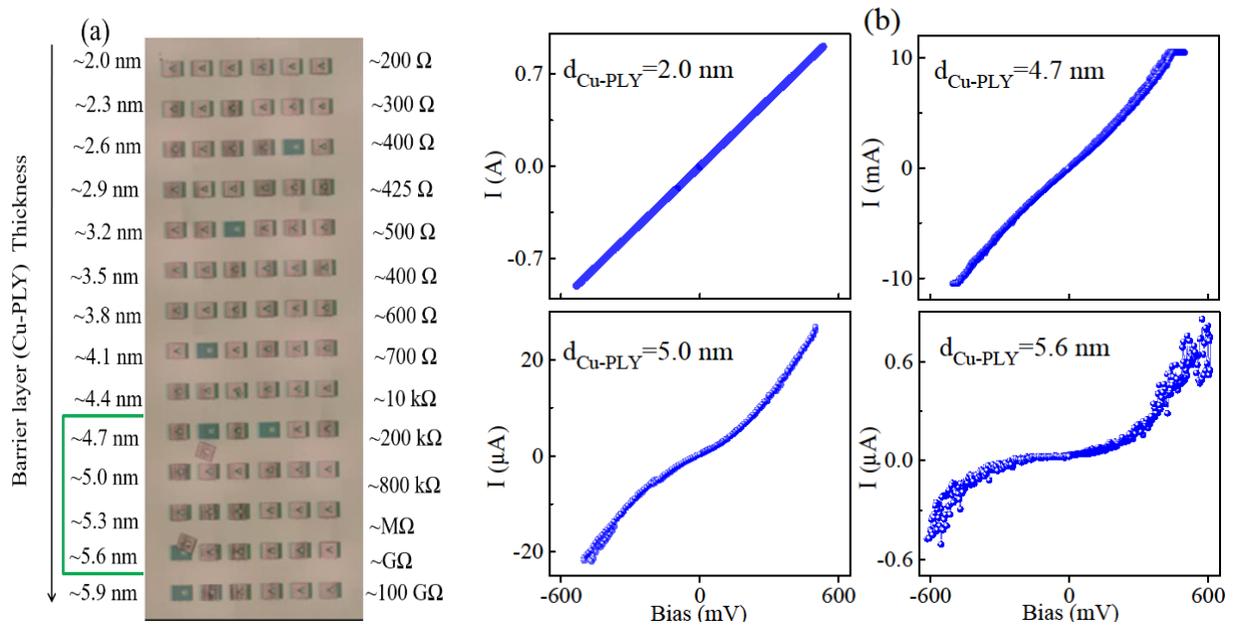

Figure S4: (a) optical image of array of Cu-PLY based OMTJs with different barrier thickness. (b) Room temperature I-V response of OMTJs with different thickness.

In order to investigate spin transport in OMTJs, firstly we need to have OMTJ with a thin barrier layer, so we fabricate sets of wedge devices (seen from Figure. S4(b)) with different organic barrier thickness from 2 nm to 6 nm and room temperature I-V measurements carried out at $\mp$ 0.5 mV, using home-built 4-point probe cryostat setup. As seen from Figure. S4(b), I-V curves showed a transition from linear to nonlinear response with increasing Cu-PLY thickness from 2 nm to 5.9 nm, which means OMTJ devices changes from the low-resistance state to the high-resistance state. The very low resistance of the OMTJs with barrier thickness less than 2.5 nm is due to the presence of pinholes in organic layers or forming a discontinuous film at organic materials/Co interface, which makes devices short-circuit. In OMTJs with barrier thickness more than 4.4 nm, resistance increases exponentially, such that OMTJs with ~ 5.6 nm thick organic layers, show resistance in GΩ range, with almost no detectable current below 1 V. Combining these results with AFM, reveal that an optimum thickness of ~ 5 nm must be considered for further investigation.

Further Magnetic characterization of this device is carried out to examine how the adsorption of molecules is changing the property of ferromagnetic electrodes.

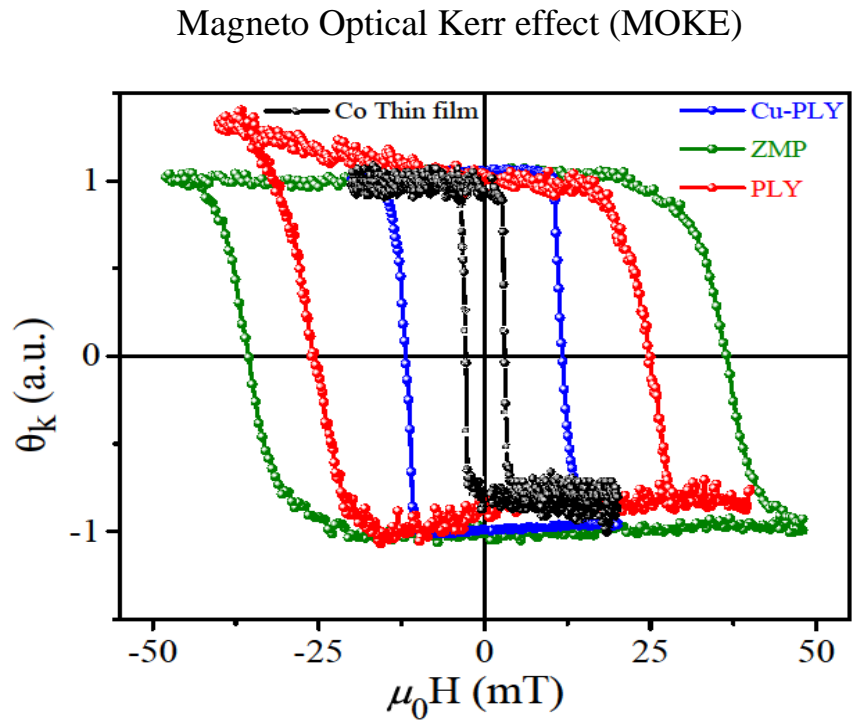

Figure S5: Room temperature MOKE measurement of OMTJs Co (8 nm)/OM (5nm)/Co (12 nm)/MgO (4 nm) with Cu-PLY(blue), PLY(red), and ZMP(green) as organic layers. Black curve represent MOKE on Co(8nm).

The magnetic hysteresis loops of Co (8 nm) thin film and OMTJ devices having stacking layers of Co (8 nm)/OM (5nm)/Co (12 nm)/MgO (4 nm), with PLY, ZMP, and Cu-PLY as an organic layer, is measured by MOKE technique at room temperature shown in Figure. S6. The Co (8 nm) (black curve) is switching at $H_c \sim 5 - 6$ mT. In the OMTJ devices as organic materials adsorbed on Co bottom layers induces different interfacial interaction (or coupling strength) between the bottom Co-electrode and the π-orbital electrons of the molecular layers which modify the surface property Co atoms and increase the coercivity and switching field. The switching field $H_c \sim 11.5$ mT, $H_c$

~ 25 mT, and $H_c$ ~ 36.5 mT respectively for Cu-PLY (blue curve), PLY(red curve), and ZMP(green curve) based OMTJs.

## Temperature dependence of Current-Voltage and Magnetoresistance characteristics of OMTJs

The current-voltage (I-V) characteristics of OMTJ devices Co (8 nm)/OM (5nm)/Co (12 nm)/MgO (4 nm), with PLY, ZMP, and Cu-PLY as organic layer, at temperature 293K and 200 K show non-ohmic behavior that confirms tunneling occurs through the organic layers. Importantly, using the tunneling model of Brinkman, Dynes, and Rowell (BDR)[5], we can fit the I-V curves to extract the values of the mean tunnel barrier height (ϕ) and thickness (d). The BDR's J(V) equation is

$$J(V) = \left(\frac{3.16 \times 10^{10} \phi^{0.5}}{d} \exp(-1.025 \times \phi^{0.5} d)\right) \left[V - \frac{A_0 \Delta \phi}{32 \phi^{1.5}} eV^2 + \frac{3 A_0^2}{128 \phi} e^2 V^3\right]$$

where J is the current density (A/cm²), $A_0 = \frac{4d\sqrt{2m_{eff}}}{3h}$ and $m_{eff}$ is the effective electron mass (kg), and Δϕ is the barrier asymmetry (V).

| Molecular layer | ϕ (eV) 200K | ϕ (eV) 293K | d (nm) AFM | d (nm) 200K | d (nm) 293K |
|---|---|---|---|---|---|
| PLY | 1.00 | 1.05 | ~ 5.0 | 3.44 | 3.34 |
| Cu-PLY | 1.01 | 0.98 | ~ 5.0 | 3.54 | 3.65 |
| ZMP | 1.13 | 2.05 | ~ 5.0 | 2.72 | 2.11 |

Table 1: Parameters extracted from J-V fitting using BDR model.

In the low bias region, eq. was used to fit the I-V curves from -500 mV to +500 mV, giving the values of the barrier thickness and height of PLY, Cu-PLY and ZMP based OMTJs at temperature 200K and 293K (table1). There is a slight difference between the obtained values from AFM and the fitted values of "d".

Among all fabricated devices, only a limited number showed nontrivial MR signal and nonlinear I-V response, simultaneously. The responses of one of optimized devices in shown in Figure S6.

The nonlinear I-V response confirms that the device is fulfilling tunneling condition as a dominating mechanism for electron/spin transport, and by using BDR's model the barrier height has low values of ϕ = 1.32 eV at 200K and ϕ = 1.10 eV at 293K. Moreover, as seen in Figure. S6(b) for OMTJ with barrier height ϕ = 1.32 eV, room temperature MR signal is weak and noisy ($MR_{avg}$ is ~ 31% ) due to the phonon contribution. We observe an increase of MR ($MR_{avg}$ is ~ 22%) and particularly a sharp peak at 3.8 mT with decreasing temperature down to 200K.

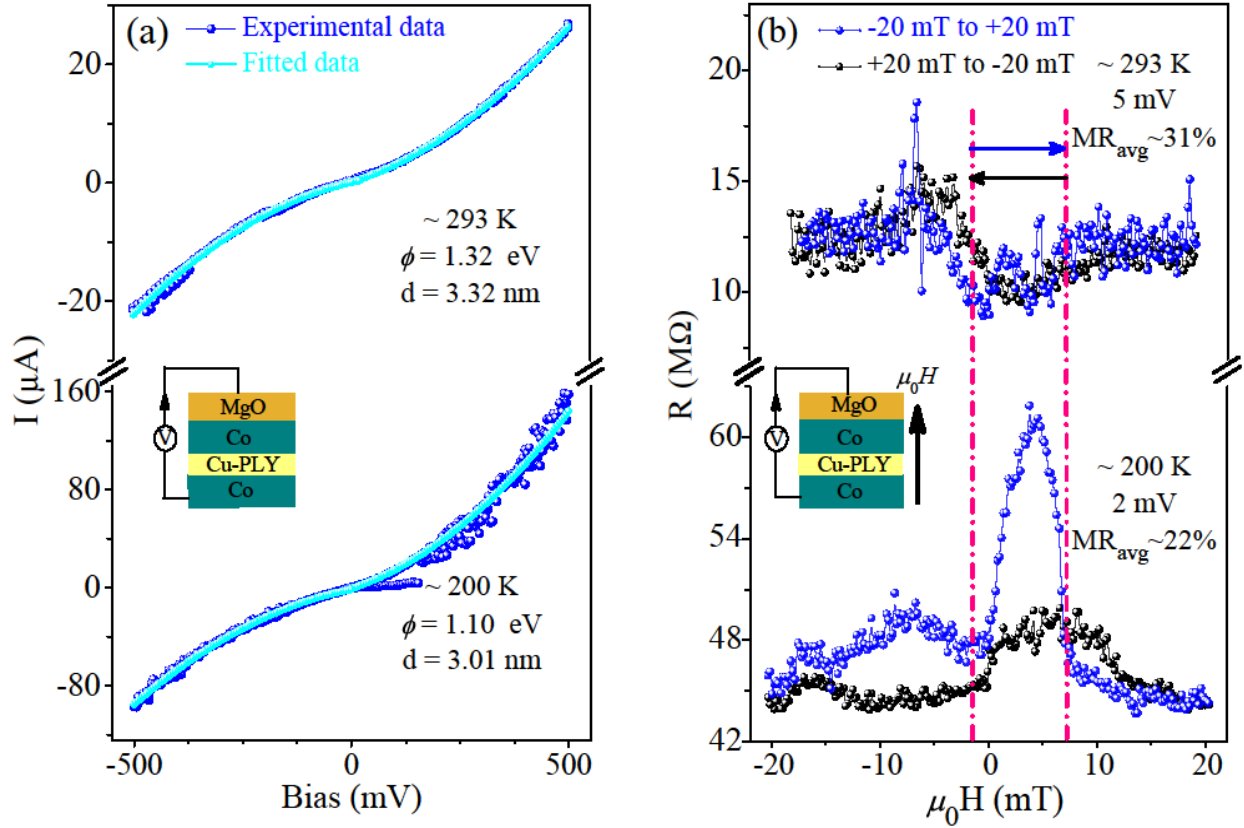

Figure S6: (a) Measured and fitted I-V characteristics and (b) Resistance variation with applied magnetic fields, for OMTJ Co/Cu-PLY/Co/MgO at temperature 293K and 200K.

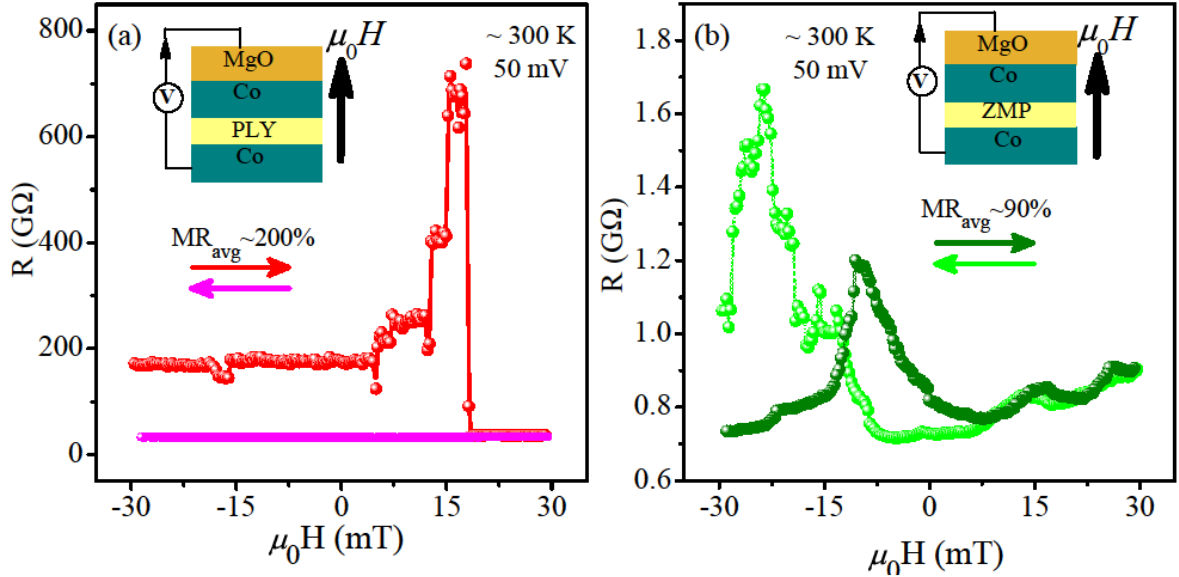

Figure S7: Magnetoresistance of PLY and ZMP based OMTJs at 50 mV. The arrows represent the direction of measurement.

As seen in Figure S7(a), MR signal is high when sweeping from the positive-to-negative magnetic field (sharp switching of resistance value red line) but when the polarity of the sweeping magnetic field is changed (positive to negative (pink line), the switching is insignificant with respectively. Figure S7(b), shows ZMP based OMTJ, MR signal is which high with $MR_{avg}$ is ∼ 90% and appears for both the magnetic field polarities sweep and we could be able to measure MR on multiple devices. This high value of $MR_{avg}$ is arises due to small junction area electron/spin mainly tunnel (weak scattering phenomena). MR measurements carried out for PLY and ZMP based OMTJ devices at room temperature. The device is highly resistive at room temperature so despite the high signal i.e. $MR_{avg}$ is ∼ 90%,, the reproducibility was less.

## Voltage Driven Resistance Switching in PLY and ZMP based OMTJs

Finally, we deal with the current–voltage (I–V) characteristics of ZMP and PLY-based OMTJs. As seen in Figure. S8, below a cutoff voltage 4V, the OMTJs are initially in an off state, which indicates the insulating nature of the devices. By increasing bias voltage, the current jumps to a high value (the device is in a low resistance state) and decreases as the reverse bias increases (the device is in a high resistance state), similar behavior has been reported in[6].

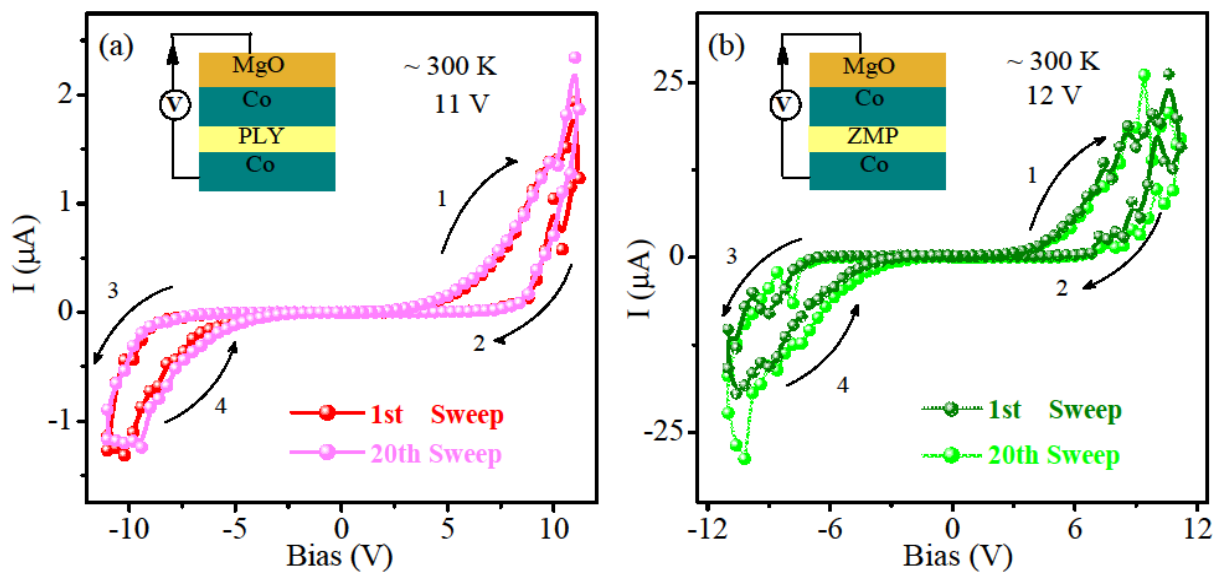

Figure S8: Memristive characteristic of (a) PLY and (b) ZMP based OMTJs at 12 V. The arrows represent the direction of measurement. Operational stability of the devices is proven by multiple (1st and 20th sweeps) I-V measurements.

This hysteretic response in the current-voltage (I-V) characteristics is a typical feature of a resistive switching device[7,8]. When compared to Cu-PLY I-V bi-stable switching response, here for PLY and ZMP the I-V hysteresis differ Figure. S9, but both have memristive like behavior with different underline mechanism[8]. The observed hysteresis in I-V traces could be explained by the presence of a molecular state which becomes charged during the voltage sweep[8]. For our devices, the mechanism is non-trivial as we are using FM electrodes separated by organic layers and the property of organic layers can be tuned by external stimuli. The resistive switching behavior could also arise due to spin transfer torque effect[9].

## Cu-PLY Crystallographic Details

A suitable single crystal of Cu(PLY)$_2$ was mounted on a glass pip. Crystallographic intensity data were collected on a SuperNova, Dual, Mo at zero, Eos diffractometer. The crystals were kept at 100 K during data collection. Atomic coordinates, isotropic and anisotropic displacement parameters of all the non-hydrogen atoms of two compounds were refined using Olex2,[10] and the structure was solved with the Superflip[11] structure solution program using Charge Flipping and refined with the ShelXL[12] refinement package using Least Squares minimization. Crystallographic data for structural analysis of Cu(PLY)$_2$ is deposited at the Cambridge Crystallographic Data

## Single Crystal Analysis

Single crystal suitable for single crystals X-ray diffraction (SCXRD) was grown by dissolving vacuum dried crystalline solid in dry dimethylformamide (DMF) at room temperature. Cu(PLY)$_2$ crystallized in monoclinic crystal system with P21/c space group. The crystal data and structure refinement for Cu(PLY)$_2$ are given in Table S2.

Table S2: Crystal data and structure refinement for Cu(PLY)$_2$.

|  | Cu-PLY |
|---|---|
| CCDC number | 2141495 |
| Empirical Formula | $C_{26}H_{14}CuO_4$ |
| Formula weight | 453.91 |
| Temperature/K | 100.00(10) |
| Crystal System | Monoclinic |
| Space group | $P2_1$/n |
| $a$ [Å] | 4.5001(6) |
| $b$ [Å] | 13.5224(17) |
| $c$ [Å] | 14.874(2) |
| $\alpha$ [$^0$] | 90.00 |
| $\beta$ [$^0$] | 93.266(13) |
| $\gamma$ [$^0$] | 90.00 |
| $V$ [Å$^3$] | 4563.26(14) |
| $Z$ | 2 |
| $\rho_{calcd}$ [gcm$^{-3}$] | 1.668 |
| $\mu$ [mm$^{-1}$] | 1.243 |
| $F$[000] | 462.0 |
| Crystal size/mm$^3$ | 0.56 × 0.42 × 0.31 |
| Radiation | MoK$\alpha$ ($\lambda$ = 0.71073) |
| 2$\Theta$ range for data collection/° | 4.08 to 55.8 |
| Index ranges | -2 ≤ h ≤ 5, -6 ≤ k ≤ 17, -18 ≤ l ≤ 18 |
| Reflections collected | 3360 |
| Independent reflections | 1961 [R$_{int}$ = 0.0343, R$_{sigma}$ = 0.0544] |
| Data/restraints/parameters | 1961/0/142 |
| Goodness-of-fit on F$^2$ | 1.107 |

The molecular unit consists of homoleptic *bis*-chelated Cu(PLY)$_2$ complex with the asymmetric unit (Figure S9) consisting of one half PLY ligand. An interesting feature of Cu(PLY)$_2$ moiety is its complete planar. All the atoms falls on the molecular plane and the torsion angle between the two PLY framework is 0º (Figure S10). The Cu-O bond distance are 1.908 and 1.912 Å that are typical of Cu(II)-O bond distance[13] and all the four oxygen atom including the Cu(II) ion fall on the same plane. The O1'-Cu1-O2 bond angle that forms chelate with PLY gives bond angle of 92.91º, slightly higher than 90º. This is compensated by the exterior O1-Cu1-O2 (non-chelated) bond angle of 87.09 º. The O1-Cu-O1' as well as O2-Cu1-O2' bond angle is perfect 180º giving rise to ideal planar conformation (*for details see* Table S3-S6).

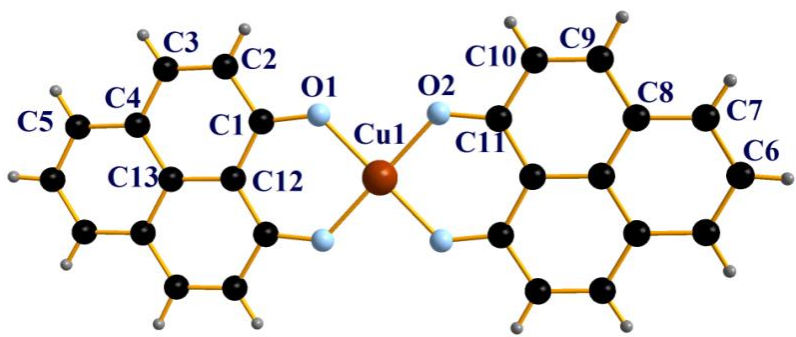

Figure S9: The ball and stick view of Cu(PLY)₂. Only the assymetric unit have been labelled for brevity.

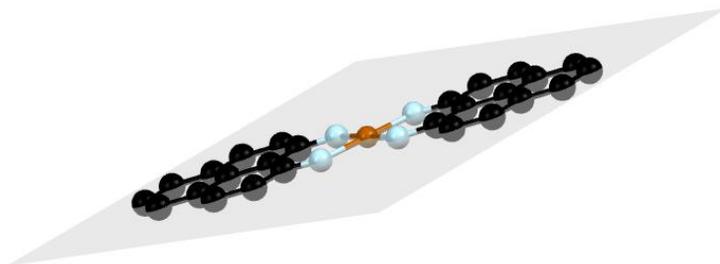

Figure S10: The ball and stick view of Cu(PLY)₂ showing the molecular plane. H-atoms are omitted for clarity.

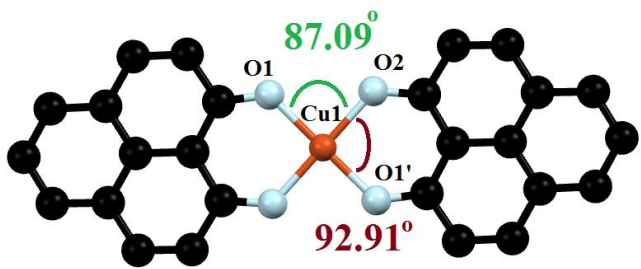

Figure S11: Variation of chelated and non-chelated O-Cu-O bond angle in Cu(PLY)₂.

Table S3: Atomic coordinates (x $10^4$) and equivalent isotropic displacement parameters (Å$^2$ x $10^3$) for Cu(PLY)₂. U(eq) is defined as one third of the trace of the orthogonalized Uij tensor.

| Atom | X | Y | z | U(eq) |
|---|---|---|---|---|
| Cu1 | 0 | 5000 | 0 | 14.95(17) |
| O1 | -566(4) | 5735.4(13) | 1074.8(12) | 17.3(4) |
| O2 | -2663(4) | 4041.2(13) | 441.7(12) | 18.0(4) |
| C1 | 618(5) | 6574.5(18) | 1315.8(17) | 13.8(5) |
| C2 | -206(5) | 6982.7(19) | 2165.2(17) | 15.5(5) |
| C3 | 933(5) | 7857.1(19) | 2482.0(17) | 14.5(5) |
| C4 | 2966(5) | 8418.4(18) | 1983.0(17) | 14.0(5) |
| C5 | 4102(5) | 9336.2(19) | 2305.3(18) | 16.2(5) |
| C6 | 6027(6) | 9878.3(19) | 1811(2) | 19.5(6) |
| C7 | 6878 (6) | 9524(2) | 981.8 (18) | 18.4 (6) |
| C8 | 5793(5) | 8618.5(19) | 638.1(17) | 15.9(5) |
| C9 | 6654(5) | 8241.3(19) | -214.2(18) | 16.9(5) |
| C10 | -5570(5) | 2631(2) | 549.5(17) | 16.7(5) |
| C11 | -3552(5) | 3228.9(18) | 58.4(17) | 14.0(5) |
| C12 | 2657(5) | 7119.7(18) | 794.6(16) | 12.5(5) |
| C13 | 3812(5) | 8049.6(18) | 1136.3(17) | 13.0(5) |

Table S4: Selected bond lengths [Å] for Cu(PLY)$_2$.*

| Atom | Atom | Length/Å |
|---|---|---|
| Cu1 | O2$^1$ | 1.9074(18) |
| Cu1 | O2 | 1.9075(18) |
| Cu1 | O1 | 1.9121(17) |
| Cu1 | O1$^1$ | 1.9121(18) |
| O1 | C1 | 1.295(3) |
| O2 | C11 | 1.291(3) |
| C1 | C2 | 1.446(3) |
| C1 | C12 | 1.438(3) |
| C2 | C3 | 1.362(4) |
| C3 | C4 | 1.429(4) |
| C4 | C5 | 1.416(3) |
| C5 | C6 | 1.379(4) |
| C6 | C7 | 1.397(4) |
| C7 | C8 | 1.403(4) |
| C8 | C9 | 1.440(4) |
| C9 | C10$^1$ | 1.360(4) |
| C9$^1$ | C10 | 1.360(4) |
| C10 | C11 | 1.445(4) |
| C11 | C12[1] | 1.432(3) |
| C11[1] | C12 | 1.432(3) |
| C12 | C13 | 1.442(3) |
| C13 | C4 | 1.426(3) |
| C13 | C8 | 1.418(4) |

*Symmetry transformations used to generate equivalent atoms: #1 -x,-y+1,-z

Table S5: Selected bond angles [°] for Cu(PLY)$_2$.*

| Atom | Atom | Atom | Angle/° |
|---|---|---|---|
| O2[1] | Cu1 | O2 | 180.0 |
| O2[1] | Cu1 | O1 | 92.91(8) |
| O2[1] | Cu1 | O1[1] | 87.09(8) |
| O2 | Cu1 | O1 | 87.09(8) |
| O2 | Cu1 | O1[1] | 92.91(8) |
| O1 | Cu1 | O1[1] | 180.0 |
| C11 | O2 | Cu1 | 127.69(16) |
| C1 | O1 | Cu1 | 127.92(16) |
| O1 | C1 | C12 | 124.4(2) |
| O1 | C1 | C2 | 117.0(2) |
| C12 | C1 | C2 | 118.6(2) |
| C8 | C13 | C12 | 120.9(2) |
| C8 | C13 | C4 | 118.6(2) |
| C4 | C13 | C12 | 120.5(2) |
| C1 | C12 | C13 | 119.0(2) |
| C11[1] | C12 | C1 | 121.9(2) |
| C11[1] | C12 | C13 | 119.0(2) |
| O2 | C11 | C12[1] | 125.1(2) |
| O2 | C11 | C10 | 116.2(2) |
| C12[1] | C11 | C10 | 118.7(2) |
| C13 | C8 | C9 | 118.5(2) |
| C7 | C8 | C13 | 119.9(2) |
| C7 | C8 | C9 | 121.6(2) |
| C5 | C6 | C7 | 120.0(2) |
| C6 | C7 | C8 | 120.9(2) |
| C13 | C4 | C3 | 118.9(2) |
| C5 | C4 | C13 | 119.7(2) |
| C5 | C4 | C3 | 121.3(2) |
| C6 | C5 | C4 | 120.8(2) |
| C2 | C3 | C4 | 121.4(2) |
| C3 | C2 | C1 | 121.6(2) |
| C9[1] | C10 | C11 | 121.5(2) |
| C10[1] | C9 | C8 | 121.4(2) |

*Symmetry transformations used to generate equivalent atoms: #1 -x,-y+1,-z

Table S6: Anisotropic displacement parameters ($Å^2$ x $10^3$) for new. The anisotropic displacement factor exponent takes the form: $-2\pi^2 [ h^2a^{*2}U^{11} + ... + 2 hka^*b^*U^{12} ]$

| Atom | $U_{11}$ | $U_{22}$ | $U_{33}$ | $U_{23}$ | $U_{13}$ | $U_{12}$ |
|---|---|---|---|---|---|---|
| Cu1 | 21.8(3) | 10.6(3) | 12.8(3) | -1.14(16) | 3.38(17) | -2.82(16) |
| O1 | 24.1(10) | 13.2(9) | 14.9(9) | -2.7(7) | 5.0(7) | -5.0(7) |
| O2 | 25.4(10) | 13.8(9) | 15.2(9) | -3.2(8) | 4.6(7) | -4.6(7) |
| C1 | 15.4(12) | 12.5(12) | 13.2(12) | 2.7(10) | -1.4(9) | 2.0(10) |
| C2 | 15.0(12) | 16.0(12) | 15.7(12) | 1.1(10) | 2.7(9) | -1.1(10) |
| C3 | 15.7(12) | 15.6(12) | 12.1(12) | 0.2(10) | 0.8(9) | 4.6(10) |
| C4 | 12.9(11) | 12.5(12) | 16.3(12) | 0.3(10) | -1.5(9) | 2.4(9) |
| C5 | 17.8(12) | 14.3(12) | 16.3(13) | -4.9(10) | -1.7(10) | 3.6(10) |
| C6 | 20.6(13) | 13.8(13) | 23.7(15) | -3.8(11) | -3.1(11) | -1.7(10) |
| C7 | 18.7(13) | 17.6(13) | 18.8(13) | 1.5(11) | -0.2(10) | -3.4(10) |
| C8 | 15.6(12) | 16.7(13) | 15.3(12) | 0.4(10) | -1.2(9) | 1.4(10) |
| C9 | 14.9(12) | 19.1(13) | 16.9(13) | 1.6(11) | 2.2(10) | -4.1(10) |
| C10 | 16.4(12) | 20.2(13) | 13.7(12) | -1.8(11) | 2.7(10) | 0.0(10) |
| C11 | 15.0(12) | 12.9(12) | 14.0(12) | -0.3(10) | -1.3(9) | 0.3(10) |
| C12 | 12.6(11) | 11.5(12) | 13.0(12) | 0.8(10) | -1.3(9) | 1.7(9) |
| C13 | 11.7(11) | 13.4(12) | 13.5(12) | -0.2(10) | -3.5(9) | 3.1(9) |

## Crystal Packing Pattern of Cu(PLY)$_2$

The existence of one-dimensional (1D) π-step stacking of phenalenyl units along the y axis is a notable structural characteristic of the Cu-PLY complex (Figure S12(a)). The packing in the x- and z-directions differs significantly from the packing in the y-direction (Figure S12(b) and S12(c)). The PLY units are connected via C5-H5⋯O1 and C5-H5⋯O2 H-bonds interactions forming a layered structure (Figure S5a). As shown in Fig. S12, PLY molecules are stacked in a sandwich pattern of face-to-face π-dimers, and adjacent planes are parallel to a distance of 3.347, 3.375 and 3.390 Å, which shows that the chain of π-stacked neighbouring phenalenyl units with intermolecular C⋯C contacts distances are shorter than the van der Waals radii (3.4 Å).

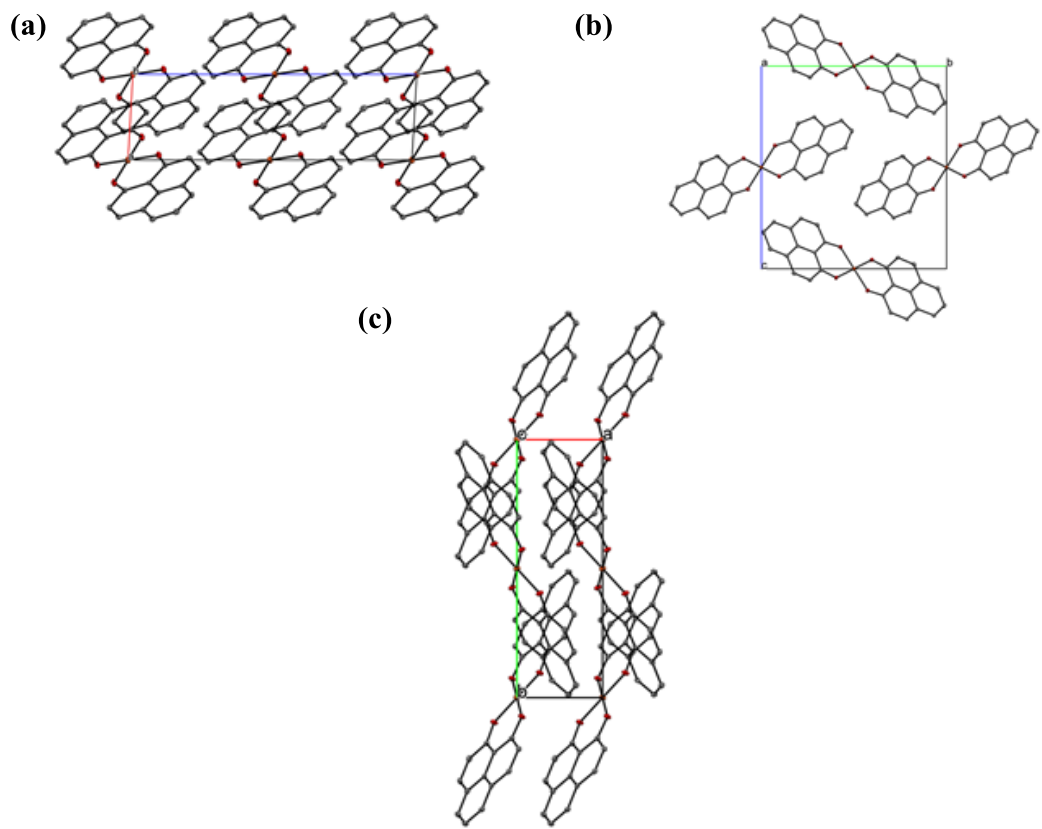

Figure S12: Packing of the molecules viewed along (a) *y*-axis, (b) *x*-axis, (c) *z*-axis. Hydrogen atoms are omitted for clarity.

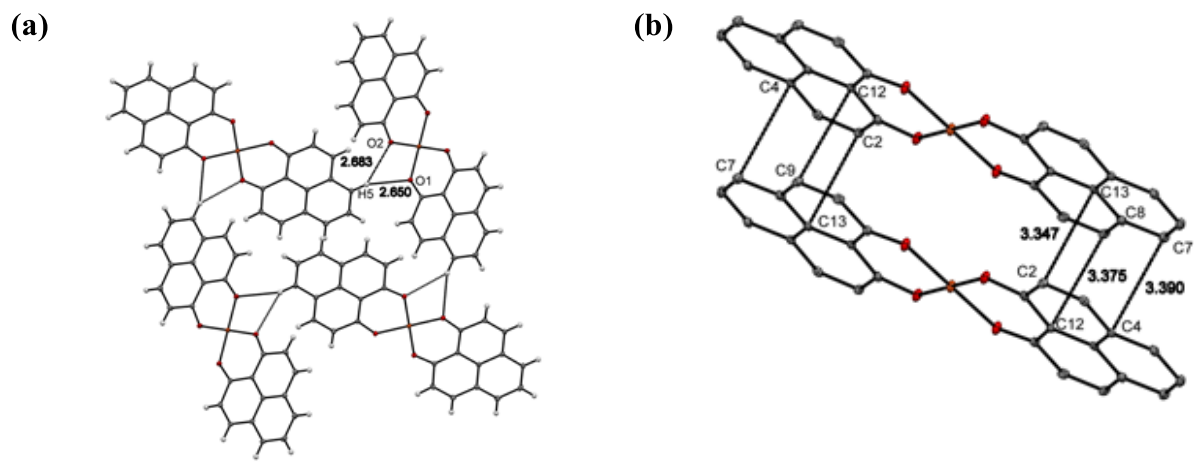

Figure S13: (a) The stacked H-bonding interaction and (b) a side view of a face-to-face overlap between the phenalenyl ligands. The phenalenyl rings are separated by an interplanar distance of 3.375, 3.347 and 3.390 Å.